\documentstyle[11pt,newpasp,twoside]{article}
\def\edcomment#1{\iffalse\marginpar{\raggedright\sl#1\/}\else\relax\fi}
\reversemarginpar

\begin{document}
\title{Four Numerical Approaches for Solving the Radiative Transfer Equation
in Magnetized White-Dwarf Atmospheres}
 \author{Stefan Jordan}
\affil{Institut f\"ur Astronomie und Astrophysik, University of T\"ubingen,
Sand 1, D-72070 T\"ubingen, Germany}
\author{Holger Schmidt}
\affil{Institut f\"ur Theoretische Physik und Astrophysik, University of Kiel,
D-24098 Kiel, Germany}
\begin{abstract}
We compare four different methods to calculate radiative transfer through a
magnetized stellar atmosphere, and apply them to the case of magnetic white
dwarfs. All methods are numerically stable enough to allow determination of
the magnetic field structure, but distinctions between faster, simplifying,
methods, and elaborate, but more CPU-time consuming, methods, can be made.
\end{abstract}
About 3\% of all known white dwarfs have strong magnetic fields between
$10^6$ and $10^9$\,Gauss (Wickramasinghe \&\ Ferrario 2000, Jordan 2001).
The detailed surface structure of the magnetic field can be inferred from
time-resolved spectro-polarimetric observations, which give integrals of the
polarized radiation over the stellar surface visible at a given rotational
phase. Because the magnetic field varies in strength and direction, theoretical
calculations for polarized radiative transfer have to be performed at
many (typically 1000) different points on the surface. In order to ensure
numerical accuracy and efficiency, we have tested four different methods from
the literature for solving the radiative transfer equations in magnetized white
dwarf atmospheres.  Note that the ALI (Accelerated \linebreak
$\Lambda$-Iteration) method is described in more detail by Deetjen et al.
in these proceedings.
\section{The Radiative Transfer Equations}
The polarization properties of light are described by the four Stokes
parameters $I$, $Q$, $V$, and $U$. $I$ is the intensity. Linear polarization
is described by  $Q$ and $U$, and circular polarization by $V$. In general,
the absorption coefficients $\kappa_l$ and $\kappa_r$ for left- and
right-handed circularly polarized light, and $\kappa_p$ for linearly polarized
radiation traveling perpendicular to the magnetic field, are not equal at a
given wavelength. $\kappa_l$, $\kappa_p$, and $\kappa_r$ correspond to
transitions where the magnetic quantum number changes by $\Delta m = -1, 0,$
and $+1$. The absorption coefficients, normalized to the Rosseland mean opacity
(\mbox{$\kappa_{\rm Ross}$}), are defined by 
$\eta_{p}=\kappa_p/\kappa_{\rm Ross}$, $\eta_{\,l}=\kappa_l/\kappa_{\rm Ross}$,
and $\eta_{r}=\kappa_r/\kappa_{\rm Ross}$. These absorption coefficients are
combined into
$\eta_{\,I}=\frac{1}{2}\eta_{p}\sin^{2}\psi+{1\over 4}(\eta_{\,l}+\eta_{r})(1+\cos^{2} \psi)$,
$\eta_{Q}= {1\over 2}\eta_{p}-{1\over 4}(\eta_{\,l} + \eta_{r})\sin^{2}\psi$,
and $\eta_{V}= {1\over 2}(\eta_{r}-\eta_{\,l})\cos\psi$. Here $\psi$ denotes
the angle between the magnetic field direction and the line of sight.
When the magneto-optical parameters \hbox{$\rho_{R}$} (Faraday rotation) and
\hbox{$\rho_{W}$} (Voigt effect) originating from spectral lines and free
electrons in a magnetic field (see Jordan et al. 1992) are taken into account,
the three radiative transport equations of Unno (1956) expand into four
equations (Beckers 1969):
\begin{equation}
\mu{\hbox{d}I\over\hbox{d}\tau} = \eta_{I}(I-B)+\eta_{Q}Q+\eta_{V}V,
\label{ri}
\end{equation}
\begin{equation}
\mu{\hbox{d}Q\over\hbox{d}\tau} = \eta_{Q}(I-B)+\eta_{I}Q+\rho_{R}U,
\label{rq}
\end{equation}
\begin{equation}
\mu{\hbox{d}U\over\hbox{d}\tau} = \rho_{R}Q+\eta_{I}U-\rho_{W}V,
\label{ru}
\end{equation}
\begin{equation}
\mu{\hbox{\hbox{d}}V\over\hbox{d}\tau} = \eta_{V}(I-B)+\rho_{W}U+\eta_{V}V.
\label{rv}
\end{equation}

The optical depth scale is $d\tau \equiv -\kappa_{\rm Ross} dz^\prime$, and 
$\mu\equiv\cos\vartheta$, where $\vartheta$ is the angle between the normal to 
the surface and the line of sight (along which the $z^\prime$ axis of the local
coordinate system on each surface element points). The $x^\prime$ axis is
defined by the direction of the projection of the magnetic field vector onto
the plane of the sky. $y\prime$ completes the right-handed coordinate system.
$B$ is the source function (the Planck formula in the case of LTE).

In the case of large Faraday rotation, radiative transfer can be described in
the framework of generalized Stokes parameters, with $I_+$ being the ordinary
mode and $I_-$ being the extraordinary mode. For the ordinary mode the electric
field vector, the magnetic field, and the direction of propagation are in one
plane; for the extraordinary mode the plane of propagation and the electric
field vector is perpendicular to the magnetic field. The phase relation between
the two modes is given by $I_c=\sqrt{I_+ I_-}\cos \delta$, and
$I_s=\sqrt{I_+ I_-}\sin \delta$, which correspond to two polarization ellipses,
the so-called normal modes.

We have used three different methods to solve the radiative transfer equations
formally: (a) The slightly modified Martin und Wickramasinghe (M\&W, 1979),
(b) an accelerated lambda iteration scheme (ALI), (c) an approximation for
large Faraday rotation (APPROX), or (d) an analytical procedure using matrix
exponential function (MATEXP).

\subsection{Algorithm by M\&W}
The solution is based on the assumption that the source function is linear in
the optical depth and that between two successive depth points the Stokes
parameters can be described by
\begin{equation}
\left(\begin{array}{c} I\\Q\\U\\V\\ \end{array}\right)=
\left(\begin{array}{c} I_a\\Q_a\\U_a\\V_a\\ \end{array}\right) +
\left(\begin{array}{c} I_b\\Q_b\\U_b\\V_b\\ \end{array}\right) \tau+
\sum_{i=4}^{4}\left(\begin{array}{c} I_{ci}\\Q_{ci}\\U_{ci}\\V_{ci}\\
\end{array}\right) \exp(a_i \tau).
\end{equation}
This assumption is inserted into the four radiative transfer equations, using
an Unno (1956) condition at the inner boundary. Then the coefficients 
$I_a,\dots V_{ci}$, as well as $a_i$, are evaluated by comparing the constant,
linear, and exponential terms on both sides of the equations. One obtains
finally a relatively simple scheme, by which the transport equations can be
solved from the inner depth points to the outside.
\subsection{ALI method}
This powerful and very flexible method is 
described in more detail by Deetjen et al. in these
proceedings.

\subsection{APPROX method}
This method uses generalized Stokes parameters
(Ramaty 1969). 
Of particular interest is the limiting case of large Faraday rotation, which
is always valid in the case of magnetic white dwarfs. This means that the
rotation between two depth points is larger than $2\pi$, and the phase relation
between the polarization ellipses $I_s$ and $I_c$ can assume arbitrary values;
$I_s=I_c=0$ on the average. Then the transfer equations decouple (Ramaty 1969):
\begin{equation}
\mu\frac{dI_{\pm}}{d\tau}=\alpha_{\pm}\left(\frac{B}{2}-I_{\pm}\right).
\end{equation}
According to V\"ath \&\ Chanmugam (1995) the formal solution (using a linear
approximation of the source function) of this equation can be written as
\begin{eqnarray*}
I_+(\tau_i)&=&\frac{1}{2}\alpha_B(1-e^{-{\alpha_+}\Delta\tau/\mu})
              \frac{1}{2}\frac{\beta_B}{\alpha_+} 
              \left[\alpha_+(\tau_i-\tau_{i+1}e^{-{\alpha_+}\Delta\tau/\mu})
                                     -(1-e^{-{\alpha_+}\Delta\tau/\mu})\right]\\
           & &+I_+(\tau_{i\pm1})e^{-{\alpha_+}\Delta\tau/\mu}\\
I_-(\tau_i)&=&\frac{1}{2}\alpha_B(1-e^{-{\alpha_-}\Delta\tau/\mu})
              \frac{1}{2}\frac{\beta_B}{\alpha_-}
              \left[\alpha_+(\tau_i-\tau_{i+1}e^{-{\alpha_-}\Delta\tau/\mu})
                                     -(1-e^{-{\alpha_-}\Delta\tau/\mu})\right]\\
           & &+I_-(\tau_{i\pm1})e^{-{\alpha_-}\Delta\tau/\mu}\\
\end{eqnarray*}

\subsection{MATEXP method}
A direct analytical way to solve the radiative transfer equations uses matrix
exponential functions. After Dittmann (1995) the radiative transfer equation
$d\vec{I}/d\tau=\mbox{\boldmath$\eta$}(\vec{I} -\vec{S})$, where
$\vec{I}\equiv(I,Q,V,U)$, can be written as
\begin{equation}
\vec{I}(0) = \exp({-\mbox{\boldmath$\eta$}\tau})
             \left[\vec{I}(\tau) -\mbox{\boldmath$\eta$}\int_\tau^0 
             \exp({-\mbox{\boldmath$\eta$}s})\,\vec{S}(s)\, ds\right]
\end{equation}
where the matrix exponential function is defined as 
$e^{\mbox{\boldmath$\eta$}}\equiv\sum_{n=0}^\infty{\mbox{\boldmath$\eta$}}^n/n!$.
The infinite series can be rewritten as $e^{\mbox{\boldmath$\eta$}\tau} = \sum\limits_{i=0}^{3} c_i(\tau) {\bf M}^i$.
The matrix ${\bf M}$ is given by ${\bf M}=\mbox{\boldmath$\eta$}-\eta_I{\bf E}$.
We assume a linear source function between successive depth points:
$S=(\alpha+\beta\tau,0,0,0)$. If we define
$x=\frac{1}{2}\sqrt{2(\eta_5^2+\rho_3^2)}$,
$y=\frac{1}{2}\sqrt{2(\eta_5^2-\rho_3^2)}$,
$z=-iy$,
$\rho_3^2=\sqrt{\eta_5^4+4(\eta_Q\rho_W +\eta_V\rho_R )^2}$,
$\eta_5=\sqrt{\eta_Q^2+\eta_V^2-\rho_R^2-\rho_W^2}$
and take into account that $y$ can be become imaginary if magneto-optical
parameters are present (a case that has been neglected by Dittmann 1995),
then the coefficients are given by:
\begin{eqnarray}
c_0&=&\frac{e^{\eta_I\tau}}{\rho_3^2}\left[
           x^2 \cos(z\tau)-y^2\cosh(x\tau)\right] \\
c_1&=&\frac{e^{\eta_I\tau}}{\rho_3^2}\left[
           \frac{x^2}{z}\sin(z\tau)-\frac{y^2}{x}\sinh(x\tau)\right] \\
c_2&=&\frac{e^{\eta_I\tau}}{\rho_3^2}\left[\cosh(x\tau)-\cos(z\tau)\right]\\
c_3&=&\frac{e^{\eta_I\tau}}{\rho_3^2}\left[
           \frac{1}{x}\sinh(x\tau)-\frac{1}{z}\sin(z\tau)\right]
\end{eqnarray}
For large magneto-optical parameters $\lim_{x\rightarrow 0}\sinh(x\tau)/x=\tau$,
and the expressions can be somewhat simplified.
\section{Conclusion}
In all tests, the resulting spectra and polarization turned out to be very
similar for the four different approaches. Small numerical instabilities in
the case of the M\&W and the MATEXP algorithms lead to small oscillations of
the Stokes parameters (particularly $Q$ and $V$) in the vicinity of spectral
lines; thsese are caused by the terms containing $\sin$, $\cos$, $\sinh$, or
$\cosh$. The ALI method, having the potential to account for NLTE and
scattering is numerically very stable. However, the APPROX method is about a
factor of 11 faster and very stable and should therefore be preferred for
practical purposes. For final fits to observations it is always possible to use
the ALI method for comparison.


\begin{references}
\reference Beckers, J.~M. 1969, Sol.Phys., 9, 372
\reference Dittmann, O. 1995, PhD thesis, Heidelberg
\reference Jordan S., 1992, A\&A, 265, 570
\reference Jordan, S. 2001, in ASP Conf. Ser. Vol. 226 (San Francisco: ASP), 269
\reference Martin, B., Wickramasinghe D., 1979, MNRAS, 189, 883
\reference Ramaty, R. 1969, ApJ, 158, 753
\reference Unno, W. 1956, PASJ, 8, 108
\reference V\"ath, H., Chanmugam, G. 1995, ApJS, 98, 295
\reference Wickramasinghe, D.~T. \& Ferrario, L. 2000, PASP, 112, 873
\end{references}
\end{document}